\documentclass[reprint,amsmath,amssymb,aps,prl]{revtex4-2}
\usepackage[utf8]{inputenc}
\usepackage[T1]{fontenc}
\usepackage{graphicx}
\usepackage{dcolumn}
\usepackage{bm}
\usepackage{xcolor}
\usepackage{lipsum} 
\usepackage{footmisc}
\usepackage{placeins}
\usepackage[colorlinks=true, linkcolor=blue, citecolor=blue, urlcolor=blue]{hyperref}
\usepackage{comment}

\newcommand{\hidecontent}[1]{} 

\begin{document}

\title{Spin-split superconductivity in spin-orbit coupled hybrid nanowires with ferromagnetic barriers}

\author{J. Zhao\textsuperscript{1}}
\author{A. Mazanik\textsuperscript{2}}
\author{D. Razmadze\textsuperscript{1}}\thanks{Current address: Microsoft Quantum, 2800 Kongens Lyngby, Denmark}
\author{Y. Liu\textsuperscript{1}}
\author{P. Krogstrup\textsuperscript{3}}
\author{F. S. Bergeret\textsuperscript{2, 4}}
\author{S. Vaitiekėnas\textsuperscript{1}}

\date{\today}

\affiliation{
\textsuperscript{1}Center for Quantum Devices, Niels Bohr Institute, University of Copenhagen, 2100 Copenhagen, Denmark
}%
\affiliation{
\textsuperscript{2}Centro de Física de Materiales Centro Mixto CSIC-UPV/EHU,\\E-20018 Donostia–San Sebastián, Spain
}%
\affiliation{
\textsuperscript{3}NNF Quantum Computing Programme, Niels Bohr Institute, University of Copenhagen, 2100 Copenhagen, Denmark
}%
\affiliation{
\textsuperscript{4}Donostia International Physics Center (DIPC),\\E-20018 Donostia–San Sebastián, Spain
}%

\begin{abstract}
We report transport studies of hybrid Josephson junctions based on semiconducting InAs nanowires with fully overlapping epitaxial ferromagnetic insulator EuS and superconducting~Al partial shells. 
Current-biased measurements reveal a hysteretic superconducting window with a sizable supercurrent near the coercive field of the ferromagnetic insulator, accompanied by multiple Andreev reflections. 
Tunneling spectroscopy shows a superconducting gap characterized by three peaks, which we attribute to tunneling between exchange-split superconductors. 
A theoretical model reproduces the observed features and indicates that spin mixing, driven by sizable spin-orbit coupling, is essential to their formation. 
Our results demonstrate proximity-induced superconductivity through a ferromagnetic insulator and establish a new platform for exploring spin-triplet pairing.
\end{abstract}

\maketitle

Hybrid structures combining superconducting (SC) and ferromagnetic insulator (FI) components enable new forms of spin-dependent superconducting transport~\cite{meservey1994spin, RevModPhys.90.041001, HEIKKILA2019100540}. 
Magnetic proximity from the FI induces spin splitting in the local density of states (LDOS) of the adjacent superconductor without requiring an external magnetic field~\cite{PhysRevLett.56.1746, PhysRevLett.61.637, tokuyasu1988proximity, hao1990spin, PhysRevMaterials.1.054402}.
Magnetic domains in the FI can further modulate superconductivity via spatially varying exchange fields~\cite{aikebaier2019superconductivity, Bergeret22sharp, diesch2018creation, maiani2025percolative}.
These effects form the basis for a range of applications, including spin valves~\cite{miao2014spin, de2018toward}, spin batteries~\cite{jeon2020giant, Silaev21enhancement}, radiation detectors~\cite{heikkila2018thermoelectric}, superconducting spintronic devices~\cite{eschrig2015spin, martinez2020interfacial}, and hybrid architectures hosting topological excitations~\cite{sau2010generic, manna2020signature, vaitiekenas2021zero}.

When used as barriers, FI can modify both the tunneling spectrum and the transport properties of the system~\cite{wei2019superconductivity, PhysRevB.100.184501, strambini2022superconducting, hijano2021coexistence}. 
In ferromagnetic Josephson junctions, spin filtering~\cite{senapati2011spin, massarotti2015macroscopic, ahmad2022coexistence} can induce $0$--$\pi$ transitions~\cite{PhysRevB.65.144502, kawabata2012spectrum, PhysRevB.74.180502, caruso2019tuning, razmadze2023supercurrent, Birge24review} and lead to spin-polarized multiple Andreev reflections (MARs)~\cite{PhysRevLett.115.116601, PhysRevB.100.060507, lu2020spin}.
Incorporating Rashba spin-orbit coupling (SOC) into these hybrid platforms mixes quasiparticle spin states~\cite{bruno1973magnetic, Meservey1975tunneling, yamashita2019proximity, Maiani2021Topological, Langbehn2021Topological, Khindanov2021topological, Escribano2021Tunable, Liu2021Electronic, Odd-frequency2018Jorge, Woods2021Electrostatic}, further modifying the behavior of the junctions~\cite{hashimoto2017tunability, bujnowski2019switchable, Minutillo2021realization} and enabling long-range spin-triplet correlations~\cite{bergeret2005odd, bergeret2014spin, bregazzi2024enhanced, tuero2024spin}. 
This interplay between spin-split superconductivity and SOC forms a rich platform for realizing and exploring unconventional Josephson effects~\cite{PhysRevB.100.104514, RevModPhys.96.021003}.

Here, we investigate the superconducting proximity effect in semiconducting InAs nanowires with thin ferromagnetic insulator EuS and superconducting Al shells.
The shells are oriented such that the Al layer is separated from the InAs core by a thin EuS barrier, similar to the planar geometry proposed in Ref.~\cite{escribano2022semiconductor}.
Transport through Josephson junctions based on these wires reveals a sizeable supercurrent, demonstrating that superconductivity can be induced through the FI barriers.
Tunneling spectroscopy on the same junctions shows a gapped spectrum with triple-peak features on both sides.
Comparison to a qualitative model suggests that the peaks arise from spin mixing induced by SOC in the proximitized semiconducting core.
The evolution of the spectrum with gate voltage indicates a tunable spin-orbit interaction in the hybrid system.

\begin{figure}[t]
    \includegraphics[width=0.48\textwidth]{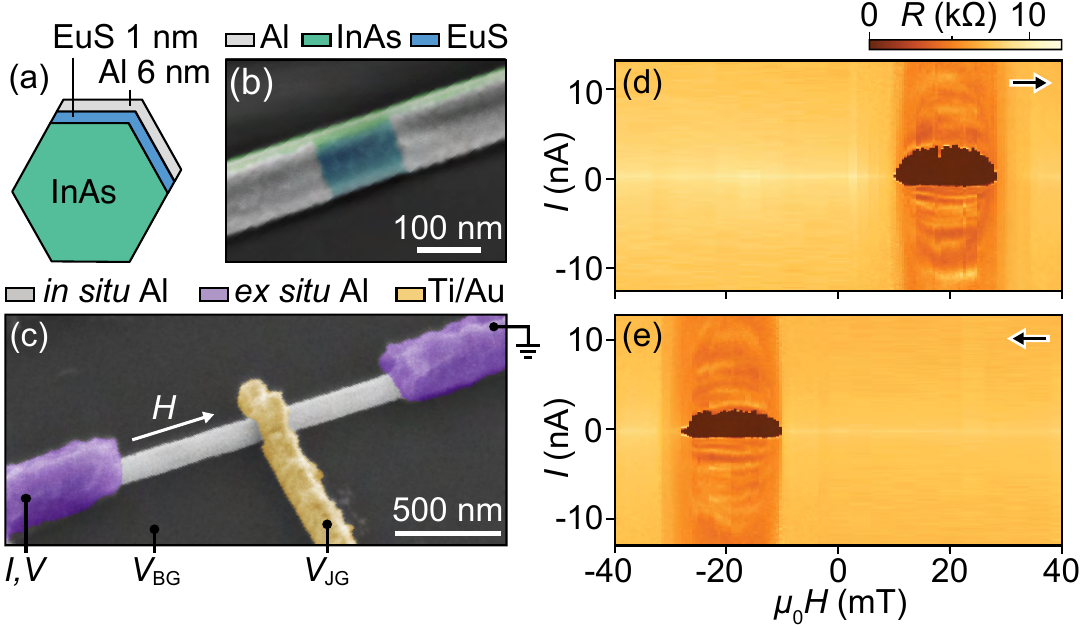}
    \caption{
    (a) Schematic of a hexagonal InAs nanowire showing EuS and Al shells fully overlapping on two facets. 
    (b) Scanning electron micrograph of nanowire after selective junction etching with enhanced false coloration of different materials.
    (c) Colorized micrograph of a representative Josephson junction device with the measurement setup.
    (d) and (e) Differential resistance, $R$, as a function of current bias, $I$, and parallel magnetic field, $H$, for junction 1 sweeping $H$ from (d) negative to positive and (e) positive to negative.
    Data were taken at $V_\mathrm{BG} = 0$ and $V_\mathrm{JG} = 0$.}
    \label{fig:1} 
\end{figure}

The measured devices are based on hexagonal InAs nanowires grown by molecular beam epitaxy (MBE)~\cite{krogstrup2015epitaxy, liu2020semiconductor}, with epitaxial ferromagnetic insulator EuS (1~nm) and Al ($6$ nm) shells deposited \textit{in situ} on the same two facets, forming a fully overlapping shell; see Fig.~\hyperref[fig:1]{1(a)}.
After the growth, the wires were transferred onto a doped Si substrate with SiO$_{\rm x}$ capping, and junctions were formed by selective etching roughly 100~nm of the Al shell using diluted TMAH in IPA, leaving the thin EuS film intact; see Fig.~\hyperref[fig:1]{1(b)}. 
The wires were contacted with \textit{ex-situ} Al leads, and individual junction top gates were metalized after atomic layer deposition of thin HfO$_{\rm x}$, enabling electrostatic control of the tunneling barrier; see Fig.~\hyperref[fig:1]{1(c)}. 
The carrier density in the wire segments on either side of the junction was tuned by the global back gate.
Four triple-hybrid junctions, denoted 1 to 4, were investigated and showed similar results. 
We report representative data from junction 1 in the main text and present supporting data from the other junctions in the Supplemental Material~\cite{Supplement}\nocite{shumeiko1997scattering,Hubler2012Observation,Virtanen22Nonlinear,Schopohl95Riccati,virtanen2020scipy,kuprianov1988influence,barone1982physics}. 
Measurements were performed using standard ac lock-in techniques in a four-probe configuration, in a dilution refrigerator with a three-axis vector magnet and a base temperature of $20$~mK.

We begin by investigating the magnetotransport properties of the junctions. 
Differential resistance, $R = dV/dI$, measured for junction 1 as a function of current bias, $I$, and the axial magnetic field, $H$, shows a well-defined superconducting window that depends on the field sweep direction; see Figs.~\hyperref[fig:1]{1(d)} and \hyperref[fig:1]{1(e)}.
Before taking the data, a magnetizing field of $\mu_0H = \pm100$~mT was applied, depending on the sweep direction.
When sweeping from negative to positive field, a finite switching current, $I_{\rm SW}$, appears around $\mu_0H=10$~mT, quickly increases to roughly 4~nA, and persists up to 30~mT, where it is suppressed again; see Fig.~\hyperref[fig:1]{1(d)}.
Sweeping in the opposite direction yields a superconducting window shifted away from $H=0$ to negative field values; see Fig.~\hyperref[fig:1]{1(e)}.
The hysteretic behavior can be understood as a consequence of magnetization reversal in the EuS layer, where the magnetic domain size near the coercive field of the FI becomes smaller than the superconducting coherence length, giving rise to reentrant superconductivity~\cite{razmadze2023supercurrent}.

\begin{figure}[t]
    \includegraphics[width=0.48\textwidth]{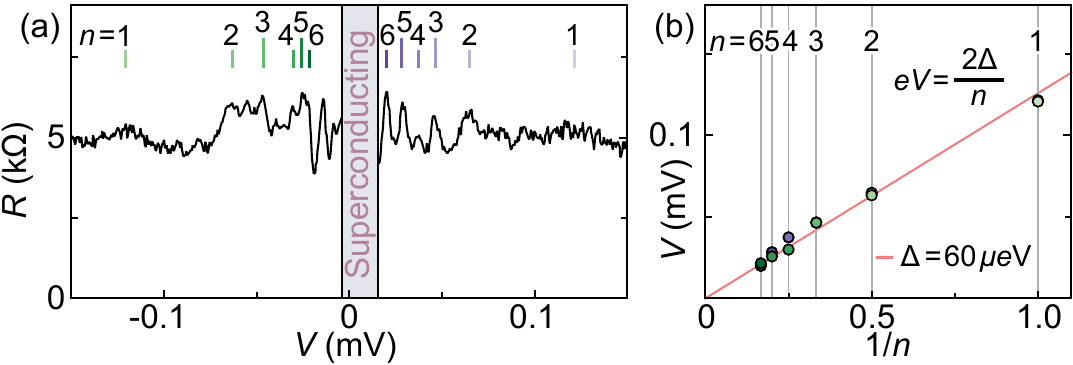}
    \caption{
    (a) Differential resistance, $R$, as function of measured voltage drop, $V$, across the junction, taken from the data in Fig.~\hyperref[fig:1]{1(e)} at $\mu_0 H = -20$~mT.
    A series of peaks in $R$ appear at finite $V$, corresponding to multiple Andreev reflections of order $n$.
    (b) Peak positions scale linearly with $1/n$.
    A linear fit to the expected relation, $V = 2\Delta/en$, yields an induced superconducting gap of $\Delta = 60~\mu$eV.} 
    \label{fig:2} 
\end{figure}

Within the superconducting window, the resistance displays oscillatory features above $I_{\rm SW}$. 
To investigate these features, we simultaneously measure $R$ and the dc voltage drop, $V$, across the junction as a function of $I$. 
A parametric plot of $R$ versus $V$, taken in the superconducting window at $\mu_0 H = -20$~mT, shows a series of peaks that appear at both voltage polarities, with increasing spacing away from $V=0$; see Fig.~\hyperref[fig:2]{2(a)}. 
We interpret these features as signatures of multiple Andreev reflections (MARs), which occur at voltages $V = 2\Delta/en$, with an integer $n$, across a junction with finite transparency~\cite{flensberg1988subharmonic, kjaergaard2017transparent}. From the measured excess current, we estimate a single-mode transparency of $\tau \approx 0.6$, consistent with the MAR interpretation (see Supplemental Material~\cite{Supplement}).
We tentatively attribute the additional peaks that appear in the superconducting window at subharmonic voltage and deviate from the standard MAR sequence to subgap structure in the leads or localized states in the junction~\cite{abay2014charge, bezuglyi2017resonant, Ridderbos2019Multiple}; these features are not analyzed further. 
The MAR peak positions scale linearly with $1/n$, yielding an induced gap of $\Delta \approx 60~\mu$eV; see Fig.~\hyperref[fig:2]{2(b)}. 
These results confirm the presence of proximity-induced gap in the semiconductor through the thin FI barrier. 

\begin{figure}[t]
    \includegraphics[width=0.48\textwidth]{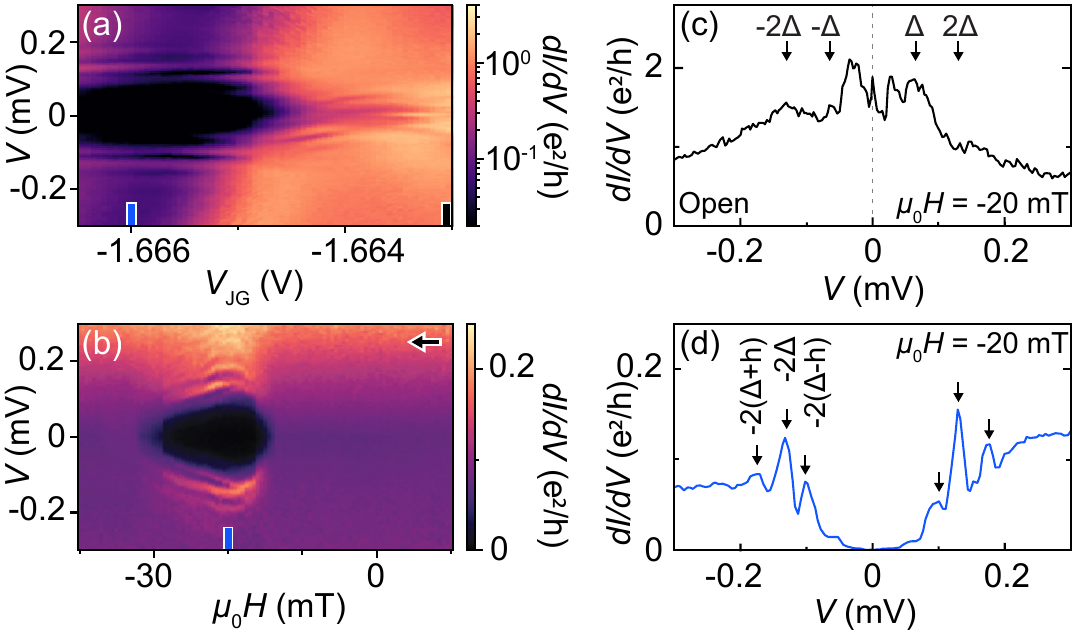}
    \caption{
    (a) Differential conductance, $dI/dV$, for junction~1 as a function of source-drain voltage bias, $V$, and junction gate voltage, $V_\mathrm{JG}$, displaying the evolution from weak-tunneling to the open regime. Data were taken at $V_\mathrm{BG} = -6.5$~V,  $\mu_0 H = -20$~mT.
    (b) Tunneling $dI/dV$ as a function of $V$ and decreasing parallel magnetic field, $H$, revealing a gapped spectrum with a triple-peak structure between  $\mu_0 H = -15$ and $-30$~mT.
    (c) Line cut from (a) in the open regime at $V_\mathrm{JG} = -1.663$~V, showing a supercurrent peak at zero bias and multiple Andreev reflection features at $V=\pm\Delta$ and $\pm2\Delta$. 
    (d) Similar to (c) but taken in the tunneling regime at $V_\mathrm{JG} = -1.666$~V, showing a gap with three well-resolved peaks.
    The central peak corresponds to $2\Delta = 130~\mu$V and is symmetrically flanked by side peaks, indicating spin splitting of $h = 40~\mu$eV.}
    \label{fig:3} 
\end{figure}

To further investigate the properties of induced superconductivity, we perform tunneling spectroscopy of the same junction in the superconducting regime at $\mu_0 H = -20$~mT, near the coercive field.
We first set the back-gate to $V_{\rm BG}=-6.5$\,V, which reduces the carrier density and electrostatically pushes hybridized wavefunction toward the superconductor interface, increasing hybridization and hardening the induced gap~\cite{vaitiekenas2018effective, antipov2018effects, mikkelsen2018hybridization, Reeg2018Metallization, moor2018electric}; similar measurements at $V_{\rm BG}=0$ are shown in Ref.~\cite{Supplement}.
We then decrease the junction-gate voltage, $V_{\rm JG}$, and record the differential conductance, $dI/dV$, as a function of voltage bias, $V$.
As $V_{\rm JG}$ is lowered, the junction transparency decreases and the conductance evolves from the open regime, with $dI/dV>e^{2}/h$, into the tunneling regime, where the normal-state $dI/dV\sim0.1\,e^{2}/h$; see Fig.~\hyperref[fig:3]{3(a)}. 
The junction enters the tunneling regime around $V_\mathrm{JG} = -1.665$~V, displaying a gapped spectrum with a gate-independent set of peaks on either side.
We note that unintentional resonances may develop at more negative $V_{\rm JG}$~\cite{ahn2021estimating}, though this regime was not studied in this work.
The magnetic-field evolution of the spectrum in the tunneling regime ($V_\mathrm{JG} = -1.666$~V) shows the superconducting gap opening around $\mu_0 H = -15$~mT, rapidly increasing and developing distinct peaks with a few millitesla, and then gradually closing around $-30$~mT; see Fig.~\hyperref[fig:3]{3(b)}.
This behavior is consistent with the superconducting window observed in the current-bias measurements [Fig.~\hyperref[fig:1]{1(e)}].

For a more quantitative analysis of the spectral features, we examine bias spectroscopy in both the open and tunneling regimes.
In the open regime ($V_\mathrm{JG} = -1.633$ V), the spectrum shows a pronounced zero-bias conductance peak, indicating supercurrent flow through the junction; see Fig.~\hyperref[fig:3]{3(c)}. 
Additional peaks appear at finite bias, including the ones at roughly 130 and 65~$\mu$V.
These are reminiscent of first- and second-order MARs and are in good agreement with $2\Delta$ and $\Delta$ extracted from current-bias measurements [Fig.~\hyperref[fig:2]{2(b)}].
In the tunneling regime ($V_\mathrm{JG} = -1.666$~V), the spectrum displays a superconducting gap defined by three peaks [Fig.~\hyperref[fig:3]{3(d)}]. The central peak is located at 130~$\mu$V,  with two additional peaks symmetrically offset by roughly 40~$\mu$V. 
We interpret this pattern as arising from spin splitting in the leads, with peak positions consistent with $2(\Delta\pm h)$, where $h$ is the effective exchange field.
We speculate that the small shoulder near $\Delta$ is a residual MAR feature. 

\begin{figure}[t]
    \includegraphics[width=0.46\textwidth]{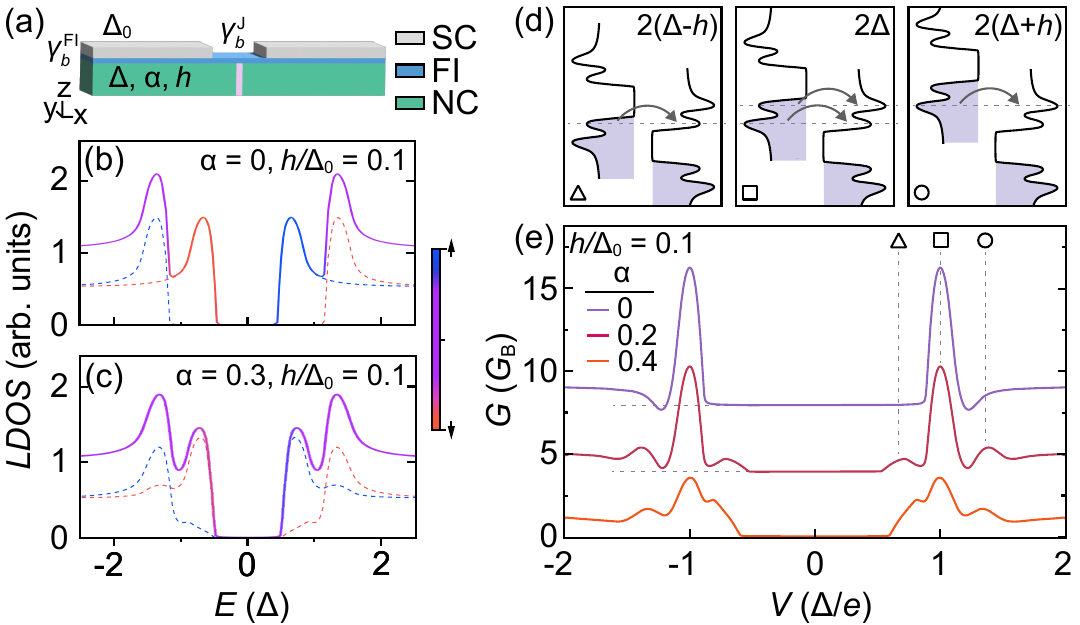}
    \caption{
    (a) Schematic of the model junction consisting of two superconducting (SC) leads with pairing potential $\Delta_0$, tunnel-coupled to normal conductor (NC) through ferromagnetic insulator (FI) barrier parametrized by $\gamma_b^{\rm FI}$.
    The NC has an induced gap $\Delta$, exchange field $h$, and dimensionless spin–orbit coupling parameter $\alpha = m\alpha_{\rm R} \xi$, where $\alpha_{\rm R}$ is the Rashba coefficient, $m$ the effective mass, and $\xi$ the superconducting coherence length.
    The junction barrier is characterized by $\gamma_b^{\rm J}$.
    (b) Local density of states, LDOS, at the junction barrier for $\alpha = 0$, showing spin-split coherence peaks due to $h$, with spin-polarized (spin-degenerate) states below (above) $\Delta$. 
    (c) Same as (b), but for $\alpha = 0.3$, with both spin components present at all energies due to spin mixing.
    (d) Schematic of resonant tunneling processes enabled by spin mixing due to finite $\alpha$. Note that for $\alpha = 0$, only the $2\Delta$ process is allowed.
    (e) Calculated differential conductance, $G$, in units of normal-state barrier conductance, $G_{\rm B}$, as a function of voltage bias, $V$, for increasing $\alpha$, showing the emergence of the triple-peak structure. Curves are offset vertically for clarity.
    Further computational details are provided in Ref.~\cite{Supplement}.
    }
    \label{fig:4_theory} 
\end{figure}

To better understand the origin of the observed triple-peak structure, we develop a simplified model based on the quasiclassical equations~\cite{larkin1986}, extended to include SOC~\cite{Bergeret13Conversion,kokkeler2025universal}.
We consider a lateral Josephson junction geometry, schematically shown in Fig.~\hyperref[fig:4_theory]{4(a)}, where each side consists of a superconducting lead coupled to a normal conductor (NC) through an FI barrier. 
In the model, the NC segments are treated as diffusive  multi-mode conductors on the coherence-length scale---an approximation justified by scattering from surface roughness and different material interfaces---which enables a quasiclassical approach~\cite{Arjoranta2016Intrinsic}.
The FI induces a spin-splitting field in both the superconducting leads and the NC, aligned along the wire axis, $x$.
Rashba spin-orbit coupling in the NC, characterized by the SOC constant $\alpha$, is assumed to originate from structural asymmetry along the $z$ direction and causes spin precession in the $x$--$z$ plane.
We note that similar effective spin–orbit interactions can also arise from spatially varying magnetic textures, such as magnetic domains or engineered micromagnets~\cite{Kjaergaard2012Majorana, Bergeret2013Singlet, Klinovaja2013Topological, bobkova2019signatures, desjardins2019synthetic}.
Assuming weak coupling between the two sides of the junction, we compute the tunneling spectra as follows.
The equilibrium Usadel equations are solved separately on each side, a gauge transformation is applied to the corresponding Green functions to incorporate the voltage bias, $V$, and the tunneling formula is used to compute the differential conductance, $G$. 
Full details of the calculation are provided in the Supplemental Material~\cite{Supplement}, and the results are summarized in Figs.~\hyperref[fig:4_theory]{4(b)--4(d)}. 

For $\alpha = 0$, the exchange field, $h$, induces spin splitting in the LDOS, resulting in a separation of the superconducting coherence peaks into spin-up and spin-down components; see Fig.~\hyperref[fig:4_theory]{4(b)}. 
Because spin is conserved for $\alpha = 0$, tunneling occurs only between equal-spin states. 
When the effective exchange field vector is identical on both sides of the junction, this leads to a single peak at $eV = 2 \Delta$ in the calculated $G(V)$~\cite{hao1990spin}; see Fig.~\hyperref[fig:4_theory]{4(e)}, purple trace.
In the superconducting state, this exchange field is responsible for singlet-triplet conversion~\cite{Bergeret13Conversion}.
When the Rashba SOC is introduced ($\alpha \neq 0$), spin is no longer conserved.
The SOC induces precession of the triplet component in opposite directions on the two sides of the junction, effectively generating magnetic inhomogeneity~\cite{Bergeret12Spinpolarized}.
As a result, a finite amount of both spin-up and spin-down components are present in the LDOS at all energies above the induced gap; see Fig.~\hyperref[fig:4_theory]{4(c)}. 
Spin mixing enables multiple resonant tunneling conditions [Fig.~\hyperref[fig:4_theory]{4(d)}], leading to three distinct conductance peaks at $eV = 2(\Delta \pm h)$ and $2\Delta$ [Fig.~\hyperref[fig:4_theory]{4(e)}, red trace], in qualitative agreement with the experimental observations.

\begin{figure}[t]
    \includegraphics[width=0.48\textwidth]{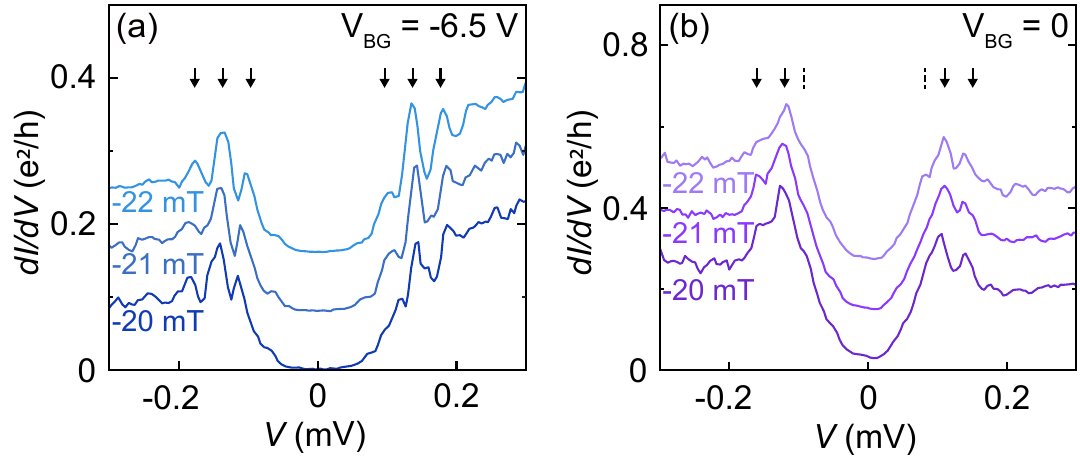}
    \caption{
    (a) Tunneling conductance, $dI/dV$, as a function of voltage bias, $V$, taken for junction 1 at several parallel magnetic field values, $H$, at back-gate voltage $\textit{V}_\mathrm{BG} = -6.5$~V, displaying a relatively hard gap with three well-resolved peaks.
    (b)~Same as (a) but taken at $\textit{V}_\mathrm{BG} = 0$, showing a softer gap and the inner peaks partially merging with the central peak, consistent with enhanced spin-orbit coupling (see main text).}
    \label{fig:5} 
\end{figure}

Increasing SOC in the model shifts the inner peaks toward higher energy and eventually causes them to merge with the central peak [Fig.~\hyperref[fig:4_theory]{4(e)}, orange trace], effectively reducing the apparent spin splitting~\cite{bruno1973magnetic}.
To test this trend experimentally, we compare tunneling conductance measurements in the superconducting window at two back-gate voltages, $V_\mathrm{BG} = -6.5$~V and  $V_\mathrm{BG} = 0$. 
At $V_\mathrm{BG} = -6.5$~V, the spectrum displays a spin-split structure with three well-resolved peaks that remain stable over a finite magnetic field range; see Fig.~\hyperref[fig:5]{5(a)}. 
Tuning to $V_\mathrm{BG} = 0$ notably shifts the inner peaks outward, reducing their separation from the central peak and blending them into a broadened central feature; see Fig.~\hyperref[fig:5]{5(b)}.

We attribute this behavior to the gate-dependent electrostatic environment, which alters the interplay between the proximity coupling and spin-orbit interaction. 
At less negative gate voltages, the hybrid wave function resides farther from the Al/EuS interface and has a larger weight in the semiconductor, resulting in a smaller and softer induced gap but enhanced SOC.
A more negative back-gate voltage reduces the charge carrier density in the semiconductor and pushes the wave function toward the interface, increasing hybridization with the superconductor and thereby hardening the induced gap while reducing the SOC~\cite{vaitiekenas2018effective, antipov2018effects, mikkelsen2018hybridization, Reeg2018Metallization, moor2018electric}.
The qualitative agreement between the model and experiment highlights the ability to tune spin-orbit coupling in this hybrid system---a key ingredient for realizing topological excitations~\cite{sau2010generic} and long-range triplet pairing~\cite{bergeret2014spin}.

In summary,  we investigated Josephson junctions based on semiconducting InAs nanowires with fully overlapping ferromagnetic insulator EuS and superconductor Al shells.
A finite supercurrent around the coercive field of EuS indicates a sizable superconducting proximity effect through the ferromagnetic barrier. 
Tunneling conductance reveals a gapped spectrum with a triple-peak structure whose evolution with gate voltage and magnetic field suggests a spin-split proximity effect shaped by spin-orbit coupling. 
A theoretical model based on quasi-classical equations links these features to spin mixing and triplet correlations. 
The ability to electrostatically tune the interplay between exchange splitting and spin-orbit interaction establishes these hybrid structures as a promising platform for engineering unconventional superconducting states.\\

\textit{Acknowledgments}---We thank S.~Abdi and S.~Telkamp for
valuable discussions, C.~S\o rensen for contributions
to materials growth, and S.~Upadhyay for nanofabrication.
Research was supported by the Danish National Research Foundation, European Innovation Council [Grant Agreement No. 101115548 (FERROMON)], research grants (Projects No. 43951 and 53097) from VILLUM FONDEN, the Spanish MCIN / AEI / 10.13039 / 501100011033 through the PID2023-148225NB-C32 project (SUNRISE) and TED2021-130292B-C42, and the EU Horizon Europe program  [Grant Agreement No. 101130224 (JOSEPHINE)].\\

\textit{Data availability}---The data used to generate the figures in this work are available in Ref.~\cite{Zhao:26}

\bibliography{Ref}
\onecolumngrid
\clearpage
\end{document}


\preprint{APS/123-QED}

\title{Supplemental Material:\\
Spin-split superconductivity in spin-orbit coupled hybrid nanowires with ferromagnetic barriers}

\author{J. Zhao\textsuperscript{1}}
\author{A. Mazanik\textsuperscript{2}}
\author{D. Razmadze\textsuperscript{1}}\thanks{Current address: Microsoft Quantum, 2800 Kongens Lyngby, Denmark}
\author{Y. Liu\textsuperscript{1}}
\author{P. Krogstrup\textsuperscript{4}}
\author{F. S. Bergeret\textsuperscript{2, 3}}
\author{S. Vaitiekėnas\textsuperscript{1}}

\affiliation{
\textsuperscript{1}Center for Quantum Devices, Niels Bohr Institute, University of Copenhagen, 2100 Copenhagen, Denmark
}%
\affiliation{
\textsuperscript{2}Centro de Física de Materiales Centro Mixto CSIC-UPV/EHU,\\E-20018 Donostia–San Sebastián, Spain
}%
\affiliation{
\textsuperscript{3}Donostia International Physics Center (DIPC),\\E-20018 Donostia–San Sebastián, Spain
}%
\affiliation{
\textsuperscript{4}NNF Quantum Computing Programme, Niels Bohr Institute, University of Copenhagen, 2100 Copenhagen, Denmark}
\maketitle

\section*{Sample Preparation}

Hexagonal InAs nanowires were grown by the vapor-liquid-solid method using molecular beam epitaxy (MBE), following the procedures described in Refs.~\cite{krogstrup2015epitaxy, liu2020semiconductor}. 
The wires have a typical length of $\sim 10$~$\mu$m and a diameter of $\sim120$ nm. 
Two of the six facets are coated with fully overlapping shells of EuS (1~nm) and Al (6 nm), deposited \textit{in situ}, in a metal deposition chamber connected to the same vacuum cluster as the III-V growth chamber.
We note that the EuS shell was grown using an established recipe~\cite{liu2020semiconductor}, though its atomic structure has not been investigated in detail.

\begin{figure}[b!]
    \includegraphics[width=0.48\textwidth]{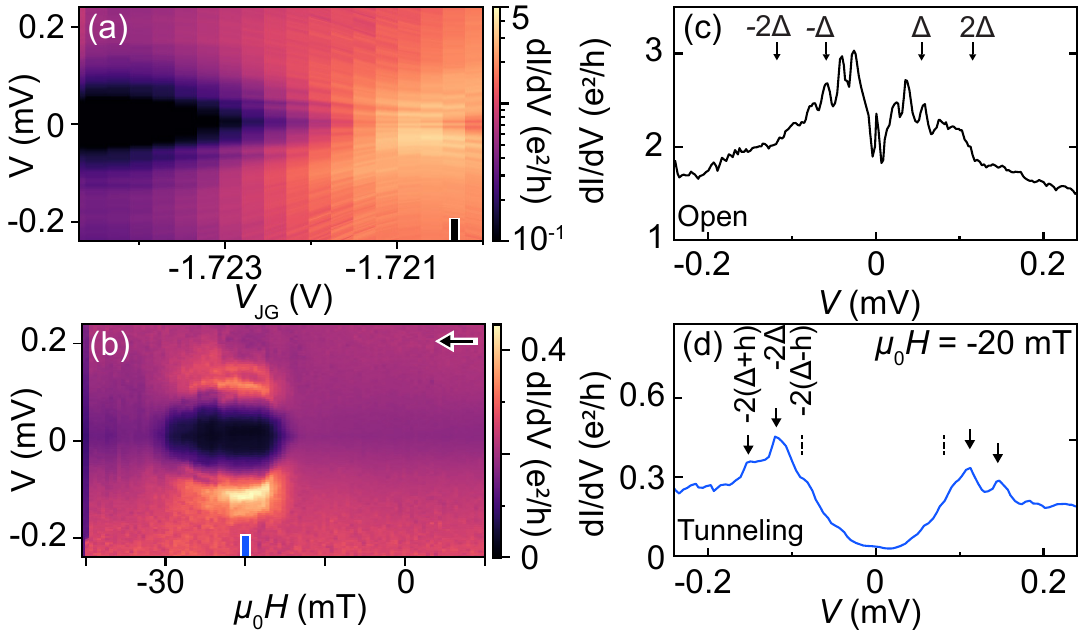}
    \caption{
    Similar to the main-text Fig.~\hyperref[fig:3]{3} but measured at $V_\mathrm{BG} = 0$. 
    (a)~Differential conductance, $dI/dV$, for junction~1 as a function of source-drain voltage bias, $V$, and junction-gate voltage, $V_\mathrm{JG}$. Data were taken at $\mu_0 H = -20$~mT. 
    (b)~Tunneling $dI/dV$ as a function of $V$,  and decreasing parallel magnetic field, $H$. 
    (c)~Line cut from (a) in the open regime at gate voltage, $V_\mathrm{JG} = -1.7205$~V. 
    (d)~Line cut taken from (b) in the tunneling regime, at $-20$ mT, showing a gap with two resolved peaks and a shoulder feature at lower energies.}
    \label{fig:S1} 
\end{figure}

Devices were fabricated using standard electron beam lithography (Elionix 7000, 100 keV). 
Individual wires were transferred from the growth substrate to a prepatterned Si chip with a 200~nm SiO$_x$ insulating layer using a micromanipulator. 
Josephson junctions of length $\sim100$~nm were defined by selective wet etching of the \textit{in-situ} Al using CSAR 62 (Allresist, AR-P 6200) 13\% resist and IPA:TMAH 0.17~N developer (Allresist, AR 300-475) 16:1
solution, optimized to minimize impact on the EuS layer.
Etching was performed for 4~min at room temperature.
The \textit{ex-situ} Al (190~nm) leads were metallized by electron-beam evaporation (AJA International Inc., Orion) following native oxide removal by Ar-ion plasma cleaning (rf ion source, 25 W, 18 mTorr, 9 min). 
Junction gates of Ti/Au (5/195 nm) were isolated from the nanowire by a 6~nm atomic layer deposited HfO$_x$ dielectric (Veeco, Savannah S100).

\begin{figure}[b!]
    \includegraphics[width=0.48\textwidth]{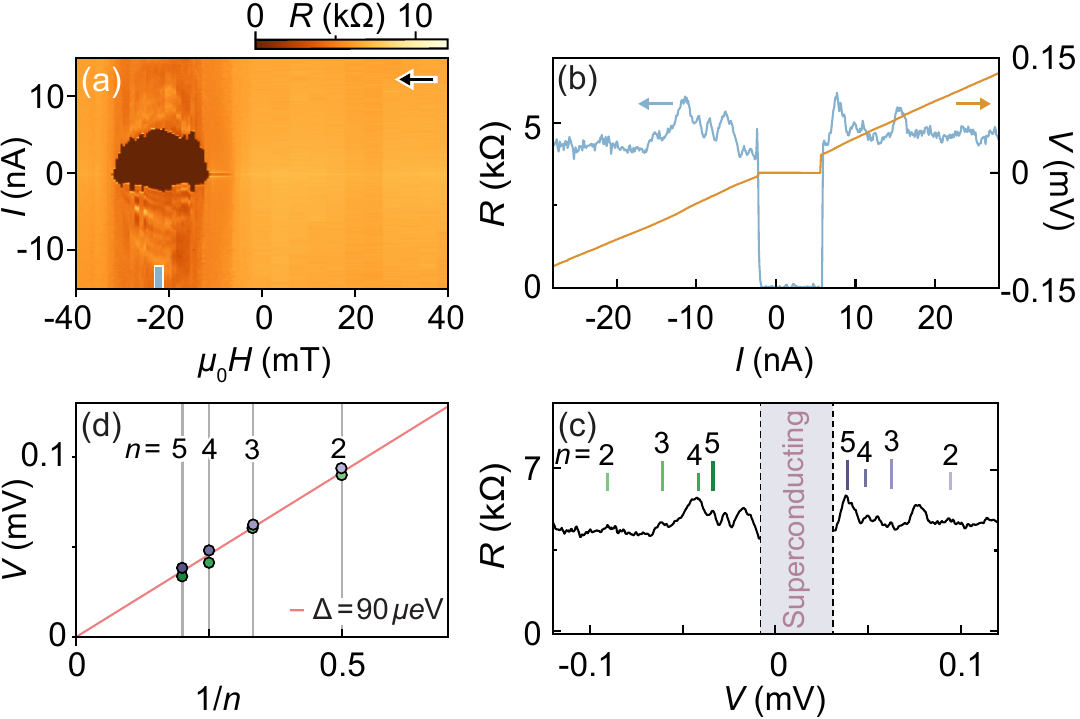}
    \caption{
    Similar to the main-text Fig.~\hyperref[fig:2]{2} but for junction~2. 
    (a)~Differential resistance, $R$, as a function of current bias, $I$, and parallel magnetic field, $H$, sweept from positive to negative. Data were taken at $V_\mathrm{BG} = 0$ and $V_\mathrm{JG} = 0$. 
    (b)~Line cut taken from (a) at $\mu_0H=-22$~mT, plotted together with measured dc voltage drop across the junction, $V$. 
    (c)~Parametric plot of data in (b) showing measured $R$ versus measured $V$. A series of peaks in $R$ appear at finite $V$, corresponding to multiple Andreev reflections of order $n$.
    (d)~Peak positions in voltage versus $1/n$. A linear fit to the expected relation, $V = 2\Delta/en$, yields an induced superconducting gap of $\Delta = 90~\mu$eV.}
    \label{fig:S2} 
\end{figure}

\section*{Measurements}

Transport measurements were performed using standard dc and low-frequency ac lock-in (Stanford Research, SR830) techniques in a cryofree dilution refrigerator (Oxford Instruments,
Triton 400) with a three axis (1,\,1,\,6)~T vector magnet and a base temperature of 20~mK. 
All measurement lines were filtered at cryogenic temperatures using rf and RC filters with a  65~kHz cutoff frequency (QDevil, QFilter).
Gate lines were additionally filtered at room temperature using home-built 16~Hz low-pass filters. 
Both ac and dc currents were transamplified using an
$I$–$V$ converter (Basel Precision Instruments, SP983) with a gain of 10$^6$, and the voltages were amplified by low-noise preamplifiers (Stanford Research, SR560) with a gain of 10$^3$. Four-probe differential resistance measurements were performed using an ac current excitation of 50~pA at 27.4~Hz.
Bias spectroscopy measurements were carried out with a 3 $\mu$V ac voltage excitation at 30.4~Hz.

\begin{figure}[t!]
    \includegraphics[width=0.48\textwidth]{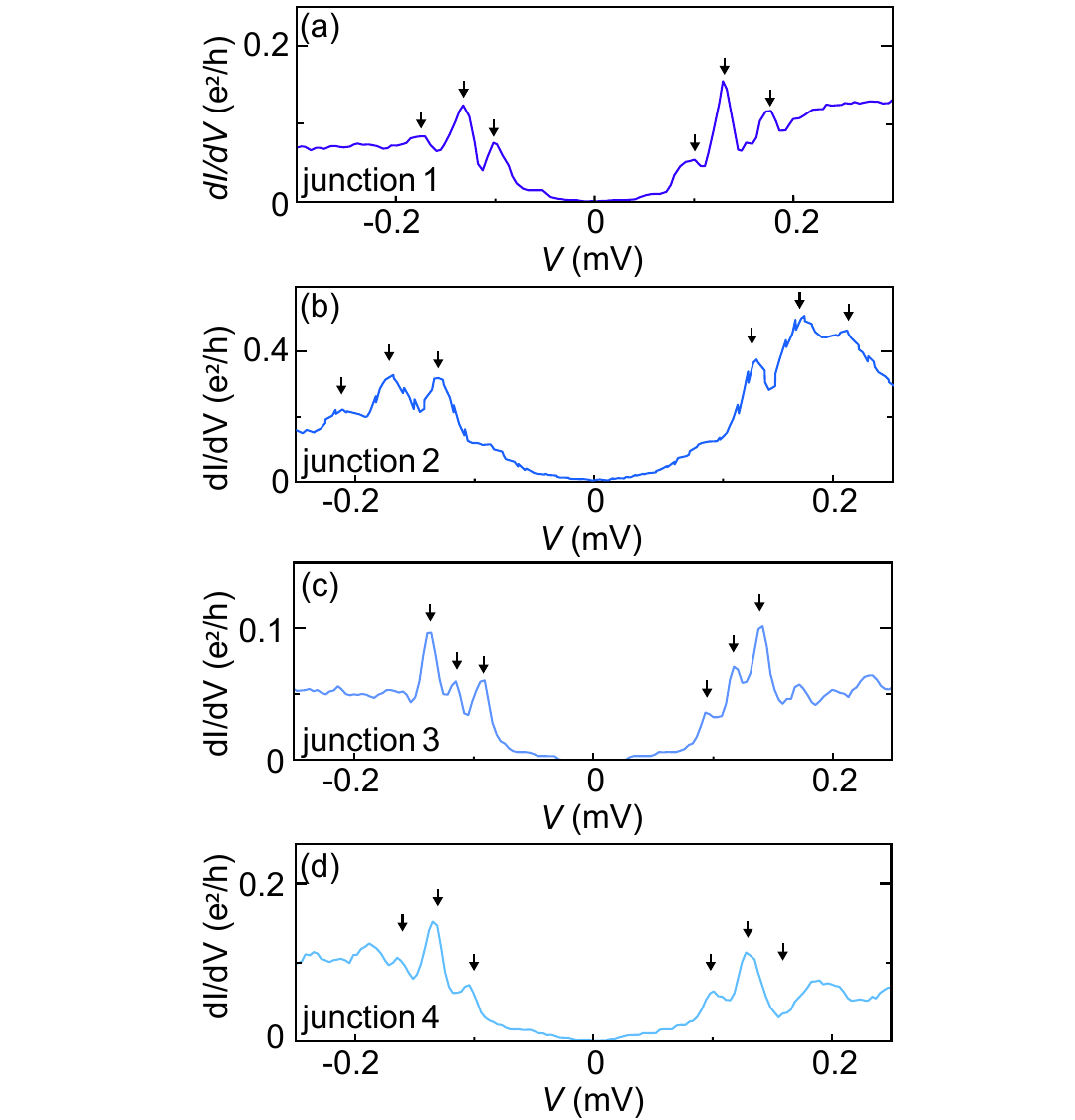}
    \caption{
    Tunneling conductance, $dI/dV$, measured for (a)~junction~1, (b) junction~2, (c) junction~3, and (d) junction~4, as a function of source-drain voltage bias, $V$, in the superconducting regime.
    All junctions show qualitatively similar spectra with three peaks at slightly different positions.
    }
    \label{fig:S3} 
\end{figure}

Four junctions, denoted 1 to 4, on two separate nanowires were measured.
Data from junction 1 are discussed in the main text, and the supplementary bias spectroscopy measurements taken at $V_\mathrm{BG} = 0$ are shown in Fig.~\hyperref[fig:S1]{S1}.
Similar measurements from other junctions are summarized in Figs.~\hyperref[fig:S2]{S2} and \hyperref[fig:S3]{S3}.

\section*{Junction Characterization}

\begin{figure}[t!]
    \includegraphics[width=0.48\textwidth]{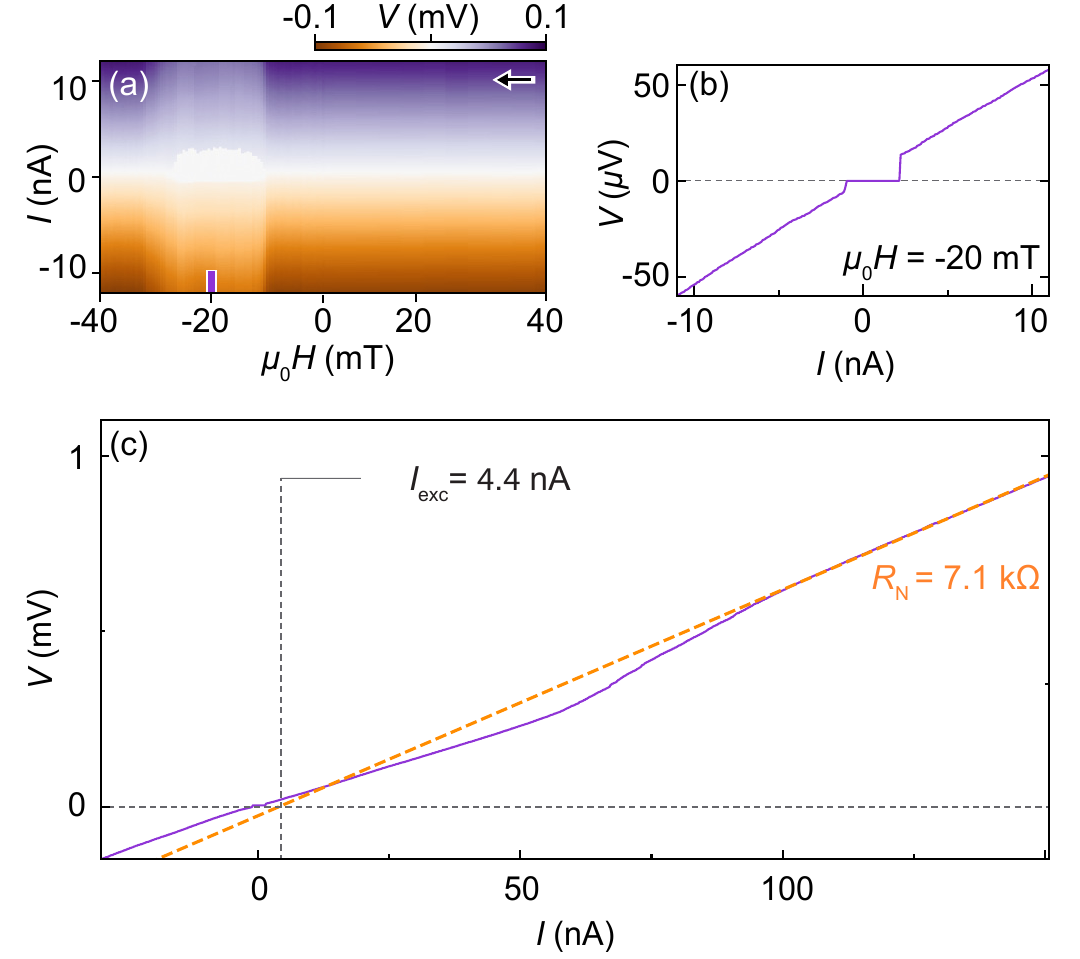}
    \caption{
    (a) Measured dc voltage, $V$, across junction~1 as a function of current bias, $I$, and parallel magnetic field, $H$, recorded simultaneously with the differential resistance data in the main-text Fig.~\hyperref[fig:1]{1(e)}. 
    (b) Line cut taken from (a) at $\mu_0 H = -20$~mT. 
    (c) Extended range of $V$ versus $I$. Extrapolation of a linear fit in the range of 95 to 145~nA indicates an excess current of $I_\mathrm{exc}$ = (4.4 $\pm$ 1.6)~nA and normal-state resistance $R_{\rm N} = 7.1$~k$\Omega$.}
    \label{fig:S4} 
\end{figure}

To characterize the triple-hybrid junctions further, we analyze their $I$–$V$ characteristics and extract the excess current, $I_\mathrm{exc}$, originating from coherent multiple Andreev reflection (MAR) processes across the superconducting weak link.
According to OTBK theory \cite{flensberg1988subharmonic}, $I_\mathrm{exc}$ depends on the dimensionless barrier strength $Z$ and is related to the transmission coefficient by $\tau = 1/(1+Z^2)$. 
The normal-state resistance is extracted from the high-bias regime, where the $I$-$V$ curve is linear, giving $R_\mathrm{N}$ = 7.1 k$\Omega$ for junction 1; see Fig. \hyperref[fig:S4]{S4}.
The extrapolated intercept at $V=0$  gives $I_\mathrm{exc}$ = (4.4 $\pm$ 1.6) nA, with the uncertainty reflecting variation across fit windows. 
Based on Ref.~\cite{flensberg1988subharmonic}, this corresponds to $Z \approx 0.8$ and a transmission coefficient $\tau \approx 0.6$.
We note that the theory of superconducting tunneling in single-mode junctions, where the effective transparency is expected to be $\tau \approx \sqrt{I_\mathrm{exc} \hbar / e \Delta}$, gives a consistent result~\cite{shumeiko1997scattering}.

\begin{figure}[t!]
    \includegraphics[width=0.48\textwidth]{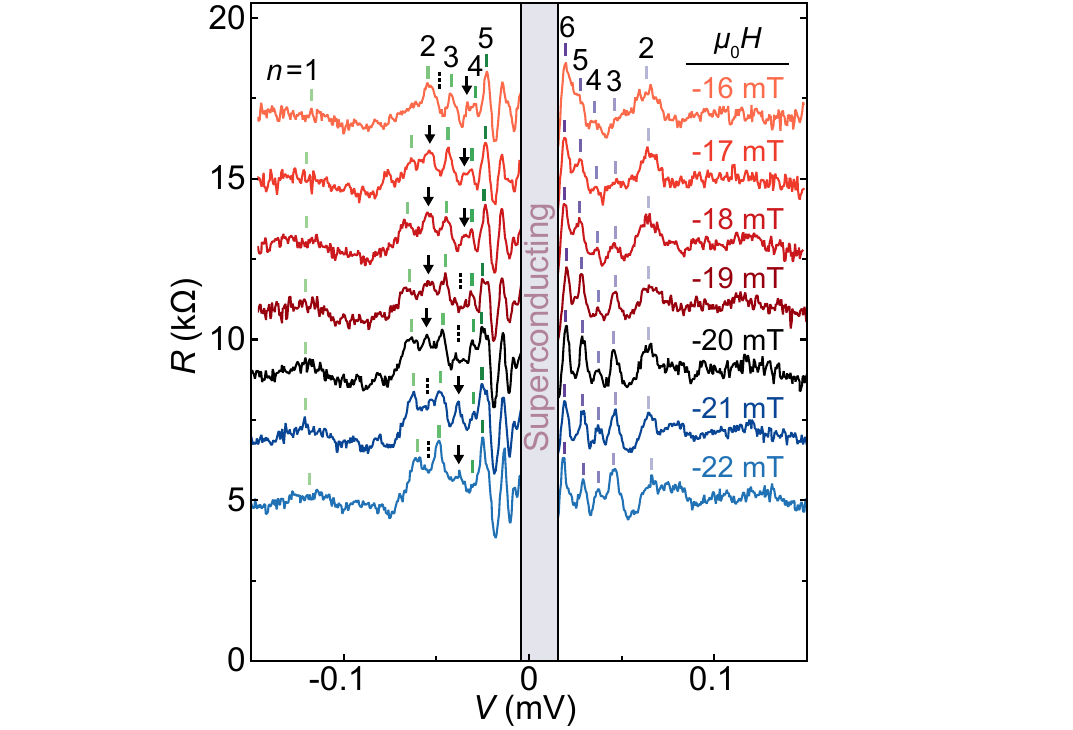}
    \caption{
    Differential resistance, $R$, as a function of measured voltage, $V$, for junction 1 at several parallel magnetic fields, $\mu_{0}H$, based on data from the main-text Fig.~1(e). 
    The line-traces were offset by 2~k$\Omega$ for clarity.
    The black trace ($\mu_{0}H=-20$ mT) is identical to the curve in Fig.~2(a). 
    Colored ticks mark the peaks corresponding to multiple Andreev reflection voltages $eV=2\Delta/n$: green for the retrapping branch and purple for the switching branch (orders $n$ indicated). 
    Additional weak features--predominantly on the retrapping branch--appear (black arrows) and disappear (dashed lines) as $\mu_{0}H$ is varied.
    }
    \label{fig:S5} 
\end{figure}

As discussed in the main text, the main MAR series is occasionally accompanied by additional, weak peaks. These peaks appear predominantly on the retrapping (negative current) branch and are largely absent on the switching branch [Fig.~\ref{fig:S5}], consistent with hysteretic non-equilibrium dynamics~\cite{Ridderbos2019Multiple}. 
Whereas the main MAR peaks evolve smoothly with magnetic field, the secondary features do not show continuous evolution with $\mu_0H$ and often disappear and reappear upon small field changes. Such behavior is consistent with contributions from subgap states in the leads or junction region whose energies are highly field sensitive~\cite{Hubler2012Observation}.
We therefore exclude these secondary features from the analysis.

\section*{Theoretical Model}
\label{sec:numerics}

\begin{figure}[t!]
    \includegraphics[width=0.48\textwidth]{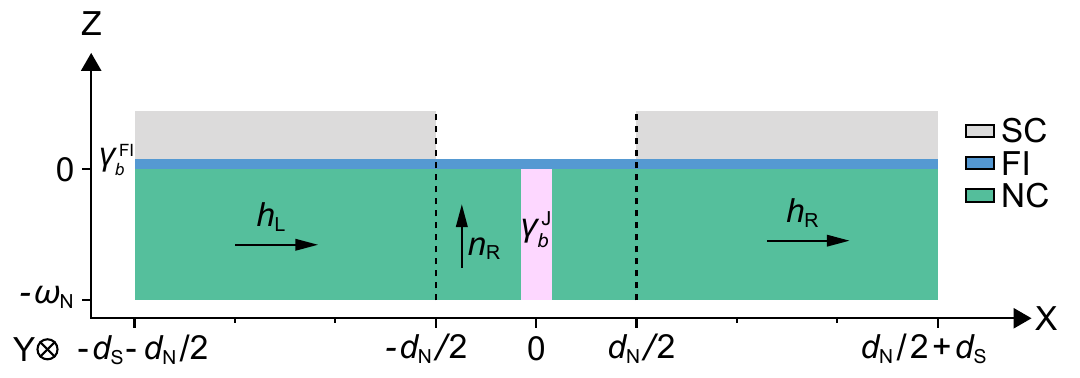}
    \caption{Detailed schematic of the theoretical setup. 
    We model Josephson junctions in the tunneling regime, with an artificial barrier at $x=0$ characterized by the boundary parameter $\gamma_b^{\rm J}$, separating left ($x < 0$) and right ($x > 0$) regions. 
    The exchange field is assumed to be the same in both the superconducting (SC) and normal conducting (NC) regions and oriented along $\vec{h}_{\rm L} = \vec{h}_{\rm R} = h\,(1,\,0,\,0)$. 
    Rashba spin-orbit coupling is present in the whole NC layer, with structural asymmetry along $\vec{n}_{\rm R} = (0,\,0,\,1)$.
    For the numerical simulations, we used $d_{\rm N} = \xi_{\rm N}$, $d_{\rm S} =  3\xi_{\rm N}$, $w_{\rm N} = 0.2\xi_{\rm N}$, $\gamma_b^{\rm J} = 10\xi_{\rm N}$, $h = 0.1\Delta_0$. 
    Here $\xi_{\rm N} = \sqrt{D/\Delta_0}$ is the coherence length in the NC layer, with the electron diffusion coefficient $D$ and the order parameter in the SC leads $\Delta_0$. The parameter $\gamma_b^{\rm FI}$ is the parameter representing the ferromagnetic insulator (FI) barrier between the SC and NC layers.}
    \label{fig:S6}
\end{figure}

We model the experimental setup in the tunneling regime using the geometry shown in main-text Fig.~\hyperref[fig:4_theory]{4(a)} and more detailed in Fig.~\ref{fig:S6}.
The system consists of two symmetric regions--left ($x<0$) and right ($x>0$)--separated by a non-magnetic tunnel barrier at $x=0$.
Each region comprises a superconducting (SC) lead connected to a normal conductor (NC) via ferromagnetic insulator (FI) barrier.
Assuming a large number of disordered conduction channels and a Fermi energy much larger than other relevant energy scales, we describe the NC regions using the quasiclassical Usadel equation with the retarded Green function, $\check{g}^{\rm R}$, represented as a matrix in spin and Nambu space as
\begin{align} \label{eq:Usadel} 
    D\tilde{\nabla}_{k}(\check{g}^{\rm R}\tilde{\nabla}_{k} \check{g}^{\rm R}) = \left[ - i \epsilon \check{\tau}_3 - i \vec{h}\cdot\hat{\bm{\sigma}}\check{\tau}_3 + i\check{\Sigma},\ \check{g}^{\rm R}\right]\; ,
\end{align}
complemented by the normalization condition $\check{g}^{\rm R}\cdot\check{g}^{\rm R} = \check{1}$.
Here, $D$ is the diffusion coefficient, $\epsilon$ is the quasiparticle energy with respect to the Fermi energy, $\hat{\bm{\sigma}}$ and $\check{\tau}_{1,2,3}$ are Pauli matrices in spin and Nambu spaces, respectively, and $\vec{h} = h\,(1,\,0,\,0)$ is the exchange field induced by magnetic proximity to the FI layer~\cite{Strambini17Revealing, Hijano21Coexistence}. 
The covariant derivatives $\check{\nabla}_k$ contain information about the spin-orbit coupling (SOC) and are defined below for two-dimensional Rashba SOC with dimensionless strength $\alpha$ and structural asymmetry along $\vec{n}_{\rm R} = (0,\ 0,\ 1)$, following Refs.~\cite{Bergeret13Conversion,Virtanen22Nonlinear}, as
\begin{equation}  \label{eq:Derivatives}
\begin{split}
    \tilde{\nabla}_x &= \partial_x -  \frac{i \alpha}{\xi}\left[ \hat{\sigma}_y,\ \right]\ ,\\
    \qquad \tilde{\nabla}_y &= \partial_y +  \frac{i \alpha}{\xi}\left[ \hat{\sigma}_x,\ \right]\ ,\\
    \qquad \tilde{\nabla}_z &= \nabla_z\; .
\end{split}
\end{equation}
The constant $\alpha$ is related to the Rashba SOC constant $\alpha_{\rm R}$ in the Hamiltonian  $\hat{H}_{\rm R} = \alpha_{\rm R} \left[\bm{\sigma}\times \vec{p}\right]\cdot\vec{n}_{\rm R}$ through the relation $\alpha = m \alpha_{\rm R} \xi$, where $m$ is the electron mass and $\xi = \sqrt{D/\Delta_0}$ is the superconducting coherence length.

The term $\check{\Sigma}$ in Eq.~\eqref{eq:Usadel} represents the self-energy that accounts for the superconducting proximity effect from the adjacent leads. 
Assuming that the normal conductor regions have a thickness $w_{\rm N}$ in the $z$ direction smaller than the characteristic length over which $\check g$ changes, we integrate Eq.~\eqref{eq:Usadel} over the thickness to obtain an effective self-energy, 
\begin{multline} \label{eq:Self-energy}
     i \check{\Sigma} =  \frac{D \check{g}^{\rm R}_{\mathrm{S}l}}{w_{\rm N} \gamma_b^{\rm J}} \left\{\theta(x + d_{\rm N}/2+d_{\rm S}) - \theta(x+d_{\rm N}/2)\right\} \\ + \frac{D \check{g}^{\rm R}_{\mathrm{S}r}}{w_{\rm N} \gamma_b^{\rm J}}\left\{\theta(x - d_{\rm N}/2) - \theta(x-d_{\rm N}/2-d_{\rm S})\right\}\;,
\end{multline}
where $\theta(x)$ is the Heaviside step function, and  $\gamma_b^{\rm J}$ is the FI barrier parameter, which is inversely proportional to the barrier conductance between SC and NC. 
The functions $\check{g}^{\rm R}_{\mathrm{S}l}$ and $\check{g}^{\rm R}_{\mathrm{S}r}$ are the Green functions of the superconducting leads, modeled as  BCS superconductors spin-split by the exchange field from the FI layer, given by
\begin{equation} \label{eq:Green_Functions_In_Leads_0}
\begin{aligned}
   \check{g}^{\rm R}_{\mathrm{S}(l,r)} &= \hat{g}^{\rm R}_{l,r} \check{\tau}_3 + \hat{f}^{\rm R}_{l,r}i\check{\tau}_2\; ,\\   
    \hat{g}^{\rm R}_{l,r} &= \frac{g^{\rm R}(\epsilon + h) + g^{\rm R}(\epsilon - h)}{2} \\ &+  \frac{g^{\rm R}(\epsilon + h) - g^{\rm R}(\epsilon - h)}{2}\hat{\sigma}_x\; ,\\
    \hat{f}^{\rm R}_{l,r} &= \frac{f^{\rm R}(\epsilon + h) + f^{\rm R}(\epsilon - h)}{2} \\ &+  \frac{f^{\rm R}(\epsilon + h) - f^{\rm R}(\epsilon - h)}{2} \hat{\sigma}_x\; ,\\
\end{aligned}
\end{equation}
with
\begin{align*}
    g^{\rm R}(x) &=  + \frac{  x \operatorname{sgn}x }{\sqrt{ (x + i\delta)^2-\vert\Delta_0\vert^2}}\ , \\
    \qquad f^{\rm R}(x) &= - \frac{  \Delta_0  \operatorname{sgn}x}{\sqrt{(x + i\delta)^2 - \vert \Delta_0 \vert^2}}\; .
\end{align*}
Here, the FI is assumed to be polarized along the $x$-axis giving $\vec{h} = h\,(1,\,0,\,0)$,  $\Delta_0$ is the superconducting order parameter in the leads, and $\delta$ is the Dynes parameter that accounts for the inelastic electron scattering.

Substituting Eqs. \eqref{eq:Derivatives}--\eqref{eq:Green_Functions_In_Leads_0} into the Usadel Eq.~\eqref{eq:Usadel}, we obtain a one-dimensional differential equation for the retarded Green function $\check{g}^{\rm R}$.
This equation is solved in the interval $x \in [-d_{\rm S} - d_{\rm N}/2,\ d_{\rm S} + d_{\rm N}/2]$, subject to boundary conditions that enforce zero-current flow at the edges of the normal regions, $\tilde{\nabla}_x \check{g}^{\rm R} = 0$ at $x = \pm(d_{\rm S} + d_{\rm N}/2)$ and at the tunnel barrier, $x = \pm0$. 
The barrier at $ x=0$ is assumed to be nearly impenetrable, allowing us to compute the tunneling conductance later based solely on the local density of states (LDOS) on either side of the barrier. 

We solve the resulting boundary value problem using the so-called Riccati parameterization of the Green function~\cite{Schopohl95Riccati}, implemented with the \texttt{scipy.integrate.solve\_bvp} routine from the \texttt{SciPy} library~\cite{virtanen2020scipy}. 
Once the retarded Green function $\check{g}^{\rm R}$ is obtained, we calculate the spin-averaged, $\nu$, and spin-resolved, $\nu_{\vec{n}}$, LDOS at $x = \pm 0$, given by
\begin{equation} \label{eq:DOS}
\begin{split}
    \frac{\nu(\epsilon)}{2\nu_0} &=  \frac{1}{4} \operatorname{Tr}\left\{ \check{\tau}_3 \check{g}^{\rm R}(\epsilon)\right\}\; , \\
    \qquad \frac{\nu_{\vec{n}} (\epsilon)}{2 \nu_0} &= \frac{1}{8}\operatorname{Tr}\left\{\left(\hat{1} + \vec{n}\cdot\hat{\bm{\sigma}}\right) \check{\tau}_3 \check{g}^{\rm R}(\epsilon)\right\} \; ,
\end{split}
\end{equation}
where $\nu_0$ is the normal-state density of states per spin at the Fermi level, and $\vec{n}$ is the direction along which the spin-resolved density of states is calculated. 
These expressions are used to generate the LDOS plots in the main-text Figs.~\hyperref[fig:4_theory]{4(b)} and \hyperref[fig:4_theory]{4(c)}.

To evaluate the tunneling current, we use the previously obtained retarded Green functions $\check{g}^{\rm R}$ at $x = \pm 0$ and construct the Keldysh Green functions at the two sides of the barrier, given by
\begin{equation}
\begin{split}
    \check{g}_{l} =\check{g}(-0) = \begin{pmatrix}
        \check{g}^{\rm R}(-0) & \check{g}^{\rm K}(-0) \\
        0 & \check{g}^{\rm A}(-0)
    \end{pmatrix}, \\ \qquad \check{g}_{r} =\check{g}(+0) = \begin{pmatrix}
        \check{g}^{\rm R}(+0) & \check{g}^{\rm K}(+0) \\
        0 & \check{g}^{\rm A}(+0)
    \end{pmatrix},
\end{split}
\end{equation}
where the advanced components are given by $\check{g}^{\rm A} = - \check{\tau}_3 {\check{g}^{{\rm R}\dagger}} \check{\tau}_3$, and the Keldysh components are parameterized through the distribution functions $\check{\varphi}_{l,r}(\epsilon) = \tanh \frac{\epsilon}{2T}\check{1}$ as $\check{g}^{\rm K} = \check{g}^{\rm R} \otimes \check{\varphi} - \check{\varphi} \otimes \check{g}^{\rm A}$, with $\otimes$ donoting a convolution in time arguments.

The voltage bias, $V$, across the junction is incorporated by applying a gauge transformation to the Green functions, which introduces time-dependent phases as
\begin{equation}
\begin{split}
    \check{g}_{l,r} \to e^{i\varphi_{l,r}(t)\check{\tau}_3/2} &\otimes \check{g}_{l,r}(t, t') \otimes e^{-i\varphi_{l,r}(t')\check{\tau}_3/2}, \\ \qquad \varphi_{l,r}(t) &= \mp eVt,
\end{split}
\end{equation}
where $\varphi_{l,r}$ are the phases of the superconducting order parameters in the left and right leads, evolving in time due to the applied potentials $\mp V/2$. 

With the voltage bias included, the Green functions $\check{g}_{l,r}$ are substituted into the Kupriyanov-Lukichev  boundary conditions \cite{kuprianov1988influence} at the interface $x = 0$ to compute the electrical current density through the junction, given by
\begin{equation} \label{eq:current}
\begin{split}
    j_x &= \frac{\pi e \nu_0 D}{4}\int \frac{d\epsilon}{2\pi} \operatorname{Tr}\left\{ \check{\tau}_3(\check{g}\check{\nabla}_x \check{g})^{\rm K}\right\} \\ &= \frac{\pi \sigma_{\rm NC} }{4e \gamma_b^{\rm J}}  \operatorname{Re} \int\frac{d\epsilon}{2\pi}\ \operatorname{Tr}\left\{ \check{\tau}_3 \left(  \check{g}^{\rm R}_l  \check{g}^{\rm K}_r + \check{g}^{\rm K}_l  \check{g}^{\rm A}_r \right) \right\},
\end{split}
\end{equation}
where $\sigma_{\rm NC} = 2 e^2 \nu_0 D$ is the normal-state conductivity of the NC layer, and $\gamma_b^{\rm J}= 2\sigma_{\rm NC} S/G_{\rm B}$ is the boundary parameter for the interface at $x = 0$, with the junction area $S$ in the $y$--$z$ plane and the normal-state conductance of the barrier $G_{\rm B}$.
As mentioned before, the barrier between the left and right regions is assumed to be nearly impenetrable, resulting in $\gamma_b^{\rm J}\gg \gamma_b^{\rm FI}$.

The total current density $j_x$ contains both equilibrium and nonequilibrium components and can be decomposed into three contributions as
\begin{equation}
    j_x = j_1 \sin 2 eV t + j_2 \cos 2eV t + j_{\rm N}(V),
\end{equation}
where $j_1$ is the conventional Josephson critical current density, $j_2$ is the amplitude of the so-called cosine term previously studied in non-magnetic junctions~\cite{barone1982physics}, and $j_{\rm N}(V)$ is the normal current density through the junction. 
Since we focus on the tunneling conductance, we discard the time-dependent terms and retain only $j_{\rm N}(V)$.
With this, the system conductance, $G$, can be finally written as 

\begin{align}
\frac{G}{G_{\rm B}} 
&= \frac{1}{G_{\rm B}} \frac{dI}{dV} \notag \\
&= \frac{d}{d (eV/\Delta)} \frac{1}{16} \operatorname{Re} 
   \int d \left( \frac{\epsilon}{\Delta} \right)\ \notag \\&\quad \operatorname{Tr}\left\{ \check{\tau}_3 \left( 
   \check{g}^{\rm R}_l  \check{g}^{\rm K}_r + \check{g}^{\rm K}_l  \check{g}^{\rm A}_r 
   \right) \right\}\; .
\label{eq:conductance}
\end{align}
This expression is used to generate the conductance curves shown in the main-text Fig.~\hyperref[fig:4_theory]{4(e)}.

\bibliography{Ref}